\documentstyle[epsfig]{mn}
\ifCUPmtlplainloaded \else

\fi
\title[Echoes from an Irradiated Disc in GRO~J1655--40]
{Echoes from an Irradiated Disc in GRO~J1655--40}
\author[R.I. Hynes et al.]{R.I.~Hynes$^1$, K.~O'Brien$^2$, Keith Horne$^2$, 
             W.~Chen$^{3,4}$, C.A.~Haswell$^{1,5}$ \\
$^1$Astronomy Centre, University of Sussex, Falmer, Brighton BN1 9QH\\
$^2$School of Physics and Astronomy, University of St. Andrews, 
    North Haugh, St. Andrews, Fife KY16 9SS\\
$^3$Department of Astronomy, University of Maryland, College Park, 
    MD 20742, USA\\
$^4$NASA/Goddard Space Flight Center, Code 661, Greenbelt, MD 20771, USA\\
$^5$Columbia Astrophysics Laboratory, Columbia University, 
    538 West 120th Street, New York, NY 10027, USA}
\begin{document}
%
%
\newcommand{\novasco}{GRO\,J1655--40}
\newcommand{\novaper}{GRO\,J0422+32}
\newcommand{\novamus}{X-ray Nova Muscae 1991}
%
%
\newcommand{\HST} {\textit{HST}}
\newcommand{\XTE} {\textit{RXTE}}
\newcommand{\RXTE}{\textit{RXTE}}
\newcommand{\GRO} {\textit{GRO}}
%
%
\newcommand{\HI}   {H\,\textsc{i}}
\newcommand{\HII}  {H\,\textsc{ii}}
\newcommand{\HeI}  {He\,\textsc{i}}
\newcommand{\HeII} {He\,\textsc{ii}}
\newcommand{\HeIII}{He\,\textsc{iii}}
%
%
\newcommand{\EBV}{E(B-V)}
\newcommand{\Rv} {R_{\rm V}}
\newcommand{\Av} {A_{\rm V}}
%
%
\newcommand{\lam}   {$\lambda$}
\newcommand{\lamlam}{$\lambda\lambda$}
\maketitle
%
%
\newcommand{\comm}[1]{\textit{[#1]}}
%
%
\begin{abstract}
We demonstrate correlated rapid variability between the optical/UV and
X-ray emission for the first time in a soft X-ray transient, \novasco:
\HST\ light curves show similar features to those seen by \XTE, but
with mean delay of 10--20\,s.  We interpret the correlations as due to
reprocessing of X-rays into optical and UV emission, with a delay due
to finite light travel time, and thus perform echo mapping of the
system.  The time-delay distribution has a mean of $14.6\pm1.4$\,s and
dispersion (i.e.\ the standard deviation of the distribution) of
$10.5\pm1.9$\,s at binary phase 0.4.  Hence we identify the
reprocessing region as the accretion disc rather than the mass donor
star.
\end{abstract}
%
%
%
%
\section{Introduction}
Soft X-ray transients (SXTs), also referred to as X-ray novae,
\cite{TS96} are low-mass X-ray binaries (LMXBs) in which long periods
of quiescence, typically decades, are punctuated by very dramatic
X-ray and optical outbursts, often accompanied by radio activity as
well.  It is commonly held that the optical emission seen during
outburst arises from reprocessing of X-rays by the accretion disc and/or
secondary star (see King \& Ritter 1998 and references therein).  It
is then natural to look for correlated X-ray/optical variability with
a view to performing echo mapping of the system, a technique which has
had great success in the study of active galactic nuclei \cite{P93}.
Such correlated variability has been seen with low time resolution in
the persistent LMXB Sco~X-1 (Ilovaisky et al.\ 1980, Petro et al.\
1981) and reprocessed optical pulsations in Her~X-1 have been used to
estimate the system parameters \cite{MN76}, but echo mapping has yet
to be fully applied to an X-ray binary.

The SXT \novasco\ was discovered in 1994 July when the Burst and
Transient Source Experiment (BATSE) on \GRO\ observed it in outburst
at a level of 1.1\,Crab in the 20--200\,keV energy band \cite{Ha95}.
After a period of apparent quiescence from late 1995 to early 1996,
\novasco\ went into outburst again in late 1996 April \cite{R96}, and
remained active until 1997 August.  During the early stages of this
outburst we carried out a series of simultaneous \HST\ and \XTE\
visits.  One of the primary goals of this project was to search for
correlated variability in the two wavebands.  The long-term evolution
of the outburst argued against significant reprocessing, as the
seemingly {\em anticorrelated} optical and X-ray fluxes observed
\cite{Hy98} are not to be expected if the optical flux is reprocessed
X-rays.  Nonetheless, significant short term correlations were
detected.  In this paper we analyse these and use them to constrain
the system geometry during outburst.  We will perform a more
comprehensive analysis of other aspects of the variability observed
during the outburst in a subsequent paper.
%
%
\section{Observations and data reduction}
\subsection{\HST}
Our \HST\ observations took place between 1996 May 14 and July 22
using the Faint Object Spectrograph (FOS) in RAPID mode \cite{K95}.
The full log is presented in Hynes et al.\ \shortcite{Hy98}.  In this
work we focus on the most promising set of observations taken on June
8 with the PRISM and either blue (PRISM/BL) or red (PRISM/RD)
sensitive detectors.  This configuration delivered a series of time
resolved spectra, termed groups, (useful spectral range
$\sim2000-9000$\,\AA) from which we constructed 2--3\,s light curves.
Over most of the spectral range of the PRISMS, the spectral resolution
is very poor, hence we can only study continuum variations.

We extracted the light curves from the count rates provided by STScI,
performing background subtraction by hand.  This is necessary as the
standard background model is known to underestimate the backgrounds by
up to 30 per cent \cite{K95}.  We rescaled the standard background
models to match unexposed regions on a group-by-group basis and
subtracted this to isolate source counts.  We then integrated the
counts over the source spectrum to obtain light curves.  The effective
bandpass can be defined by the FWHM of the {\em count rate} spectra:
3100--4800\,\AA\ for PRISM/BL and 3800--7400\,\AA\ for PRISM/RD.

A final subtlety involves the start times of the groups.  As discussed
by Christensen et al.\ \shortcite{C97}, FOS RAPID mode may produce
groups unevenly spaced in time: the ``too rapid RAPID'' problem.
Because of this the standard data products contain an uncertainty in
the start times of individual groups of -0.255\,s/+0.125\,s.  We
therefore had our start times recalculated using the
\textsc{rapid\_times} program at STScI which reduces the {\em
relative} uncertainty between groups to $<1$\, $\mu$s.  There is still
an unavoidable 0.255s uncertainty in the {\em absolute} start time
of each exposure (i.e.\ the zero point of each series of groups.)
\subsection{\XTE\ data}
The \RXTE/PCA data included in this paper were obtained on 1996 June
and contains 4 segments of exposure about 3.5\,ks each.  The original
data were taken with two standard EDS modes and three additional modes
for high time-resolution (up to 62\,$\mu$s) studies. The standard-1
mode has a 0.0125\,s time resolution and a single band covering the
entire 256 energy channels.  Since the highest time resolution from
the HST data is only about 2--3\,s, the light curves for the
correlation study performed in this paper are extracted from the
standard-1 mode data in 1\,s time bins using the {\rm saextrct} task
in the \textsc{ftools} software package. The average count rate of
\novasco\ during the above mentioned observing period is more than
$2\times10^4$\,counts\,s$^{-1}$ of ``good events'' (i.e., after 80\%
of internal background events are rejected by the anticoincidence
logic), thus no background ($\le35$ counts s$^{-1}$) subtraction
procedure is necessary.

The relative timing accuracy of the \RXTE\ data is limited only by the
stability of the spacecraft clock which is good to about 1\,$\mu$s or
less.  The absolute timing accuracy, however, is also limited by
uncertainties in the ground clock at the White Sands station and other
complications. For data taken before April 29, 1997, the absolute
timing accuracy is estimated to be about 8\,$\mu$s which is
substantially better than that of \HST/FOS and certainly sufficient
for our correlation study.

The {\it RXTE} light curves of \novasco\ exhibit variability on
various time scales. On the 1\,s or less time scale, there are
flickerings with RMS of 10\% or so. On longer time scales up to a few
hundred seconds, however, the amplitude of broad peaks and troughs
(they are not necessarily following each other) can be as high as 50\%
which are the source for the induced optical variability we detect.
%
%
\section{Comparison of light curves}
\begin{figure}
\begin{center}
\epsfig{height=2.65in,angle=90,file=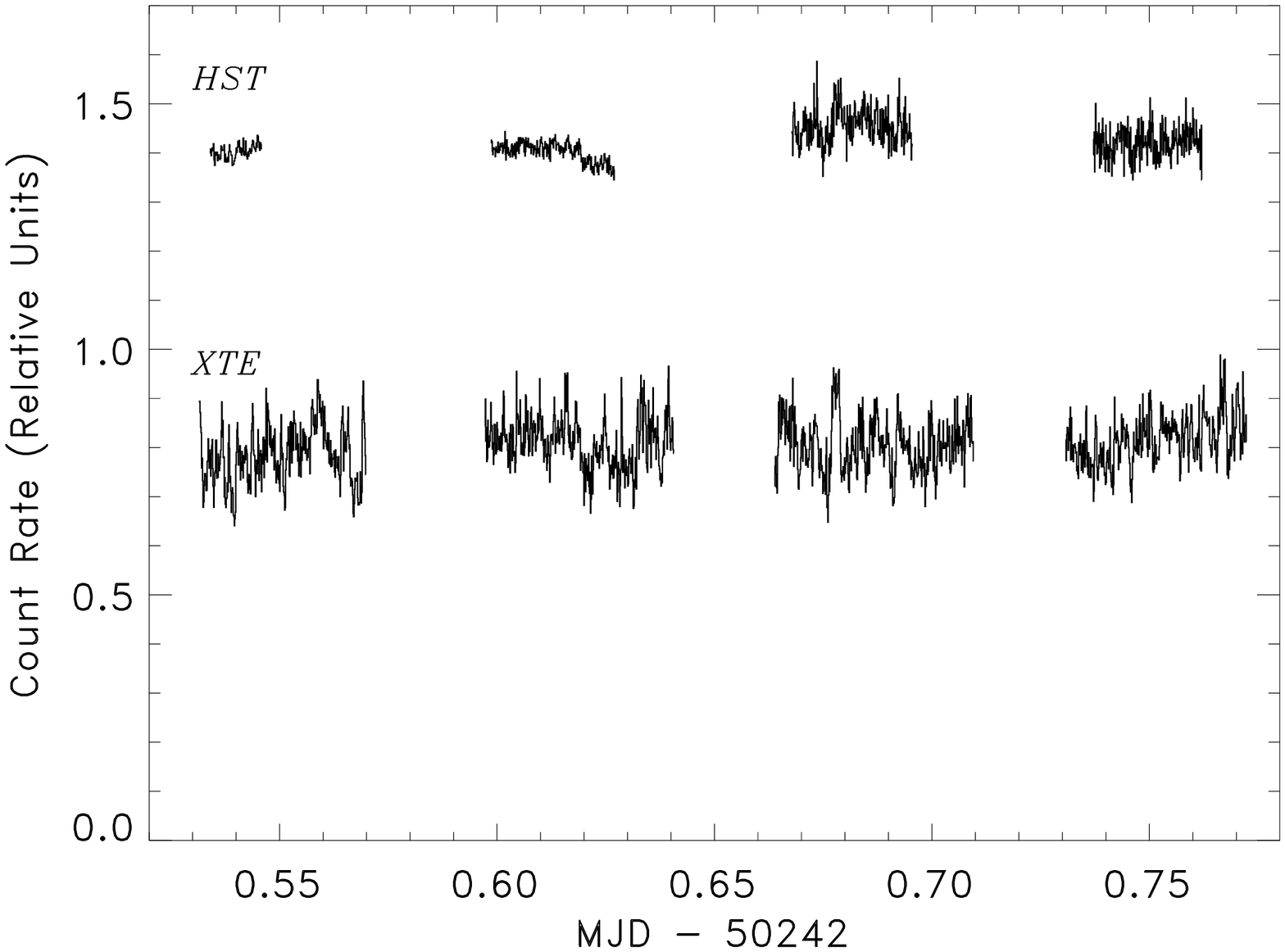}
\epsfig{height=2.65in,angle=90,file=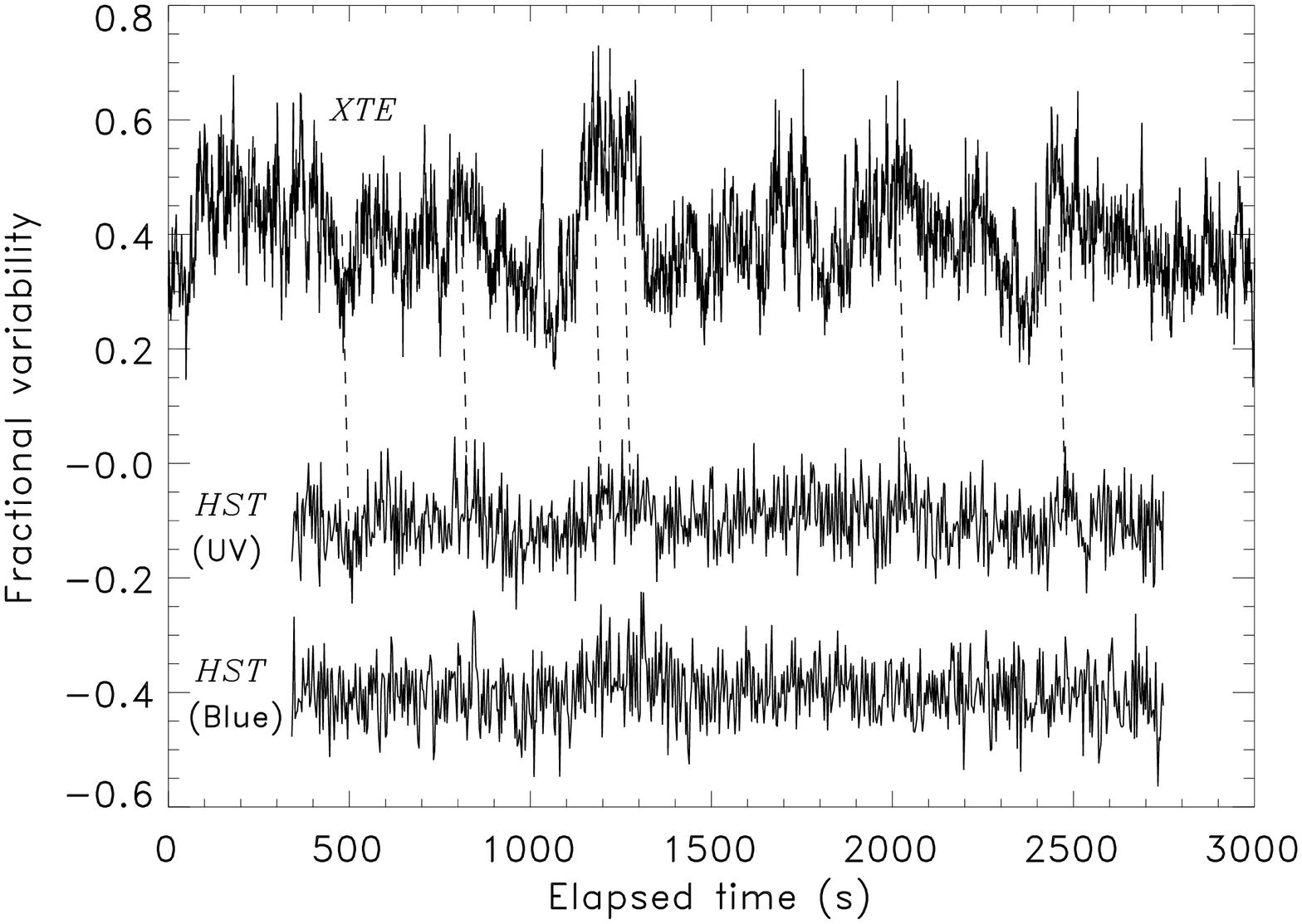}
\caption{a) \HST\ and \XTE\ light curves from 1996 June 8.  Note that
         the first two \HST\ light curves use PRISM/RD, while the
         latter two use PRISM/BL.  For clarity, the light curves have
         been rebinned to a time-resolution of $\sim$10\,s.  The
         relative count rates of \XTE\ vs PRISM/RD vs PRISM/BL have
         been rescaled but the zero point is correct.  Typical errors
         are 0.7 per cent for the \RXTE\ light curves, 1.5 per cent
         for PRISM/RD and 3.4 per cent for PRISM/BL.  b) A close up of
         the third pair of light curves from a) showing correlated
         variability and colour dependence in the \HST\ light curves.
         The zero point is arbitrary, but the vertical axis indicates
         the amplitude of variations relative to the mean count rate.
         The wavelengths compared are 2000--4000\,\AA\ (UV) and
         4000--6000\,\AA\ (blue).}
\label{MultiLCFig}
\end{center}
\end{figure}

In Fig.~\ref{MultiLCFig} we show the light curves obtained in the
June~8 visit.  Of the four visits on which both \HST\ and \XTE\
observed \novasco, this achieved the best coordination.
Fig.~\ref{MultiLCFig}\,a) shows the full data set to indicate the
extent of simultaneous coverage.  In Fig.~\ref{MultiLCFig}\,b) we show
the correlations present in the third pair of light curves.  To
illustrate wavelength dependence we show both UV (2000--4000\,\AA) and
blue (4000--6000\,\AA) light curves.  While the main feature around
1200\,s is present in both, the smaller correlated features are more
prominent, or only present at all, in the UV light curve, suggesting
that the source of variability represents a larger fraction of the
total light in the ultraviolet than in the optical.  There are also
some features which are strong in the X-ray light curve, e.g.\ at
1700\,s, but which do not appear in either \HST\ light curve.  These
conclusions are borne out by a similar close comparison of the other
light curves, and it is clear that the relation between X-ray and
optical emission is complex.  It may be, for example, that the
observed X-ray variations originate from different locations, some of
which can illuminate the disc and some of which cannot.

Another example of uncorrelated variability may the apparent downward
step in the second PRISM/RD light curve in Fig.~\ref{MultiLCFig}\,a).
There may also be a step in the X-ray lightcurve, but the two do not
match well, and the optical light curve could not be reproduced as a
convolution of the X-ray light curve with a Gaussian transfer function
(see Sect.~\ref{GaussianSection}).  The pronounced step in the optical
light curve also results in a strong autocorrelation, which leads to a
very broad peak in the cross-correlation function (see
Sect.~\ref{CCFSection}).  We therefore truncated this light curve just
before this point to simplify the analysis in the subsequent sections.

%
%
\section{Analysis}
\subsection{Cross correlations}
\label{CCFSection}
We begin by performing a cross-correlation analysis.  This will
identify correlations and reveal the mean lag between X-ray and
optical variability.  The technique is commonly used in the study of
correlated variability from active galactic nuclei (AGN) where two
methodologies have been developed: the Interpolation Correlation
Function, ICF \cite{GP87}, and the Discrete Correlation Function, DCF
\cite{EK88}.  White \& Peterson \shortcite{WP94} contrasted the
relative merits of the two and suggested some improvements.  We have
tested both methods on our data and found no significant differences,
so choose to adopt the ICF method with one important modification.
This is that since the \RXTE\ data has a finer time resolution than
the \HST\ data, we approximately integrate the \RXTE\ light curve over
each lagged \HST\ timebin, rather than simply interpolating.

\begin{figure}
\begin{center}
\epsfig{width=2.65in,file=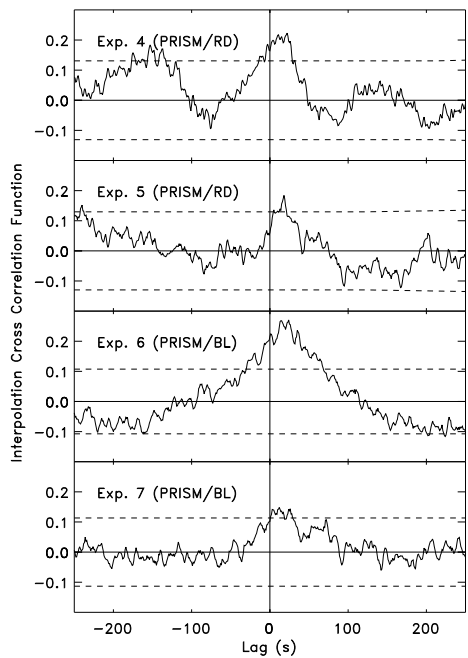}
\caption{ICFs for June~8 data.  Dashed lines show $3\sigma$ confidence
limits expected for uncorrelated variability.  All four plots show
features at $\sim20$\,s which are significant at the $3\sigma$ level,
although only Exp.\ 6 can be considered more than marginal at this level.}
\label{MultiICFFig}
\end{center}
\end{figure}

We show a section of the resulting ICFs from the June~8 visit, centred
at zero lag, in Fig.~\ref{MultiICFFig}.  All four show roughly
coincident peaks, significant at the $3\sigma$ level; if the data is
combined to yield a single ICF, then the significance of the combined
peak is $8\sigma$.  The ICF peaks occur at lags in the range
12--24\,s.  While other peaks are seen in individual ICFs, since they
are not repeated in more than one pair of light curves, we do not
judge them significant.

\subsection{Fitting transfer functions}
\label{GaussianSection}
To characterise the distribution of time delays present between the
\RXTE\ and \HST\ lightcurves, we fitted parameterised model transfer
functions.  In this modelling we predict the \HST\ light curve by
convolving the observed \RXTE\ lightcurve with a Gaussian transfer
function (i.e.\ the time-delay distribution).  We then judge the
``badness-of-fit'' by calculating the $\chi^2$ over the data points in
the \HST\ lightcurve.

This model adopts the measured \RXTE\ lightcurve verbatim, thus
ignoring the statistical errors in the \RXTE\ measurements.  This is
an acceptable approximation because the signal-to-noise ratio for
detecting variations is much higher for the \RXTE\ data than for the
\HST\ data.

The Gaussian transfer function 
\[
\psi( \tau ) = \frac{\Psi}{ \sqrt{2 \pi \Delta \tau } }
        e^{\frac{1}{2}\left( \frac{\tau-\tau_0}{\Delta\tau} \right)^2}
\]
has 3 parameters: the
mean time delay $\tau_0$, the dispersion or root-mean-square time
delay (hereafter RMS delay) $\Delta \tau$, which is measure of the
width of the Gaussian, and the strength of the response, $\Psi$, which
is the area under the Gaussian.

\begin{figure}
\begin{center}
\epsfig{height=2.65in,angle=270,file=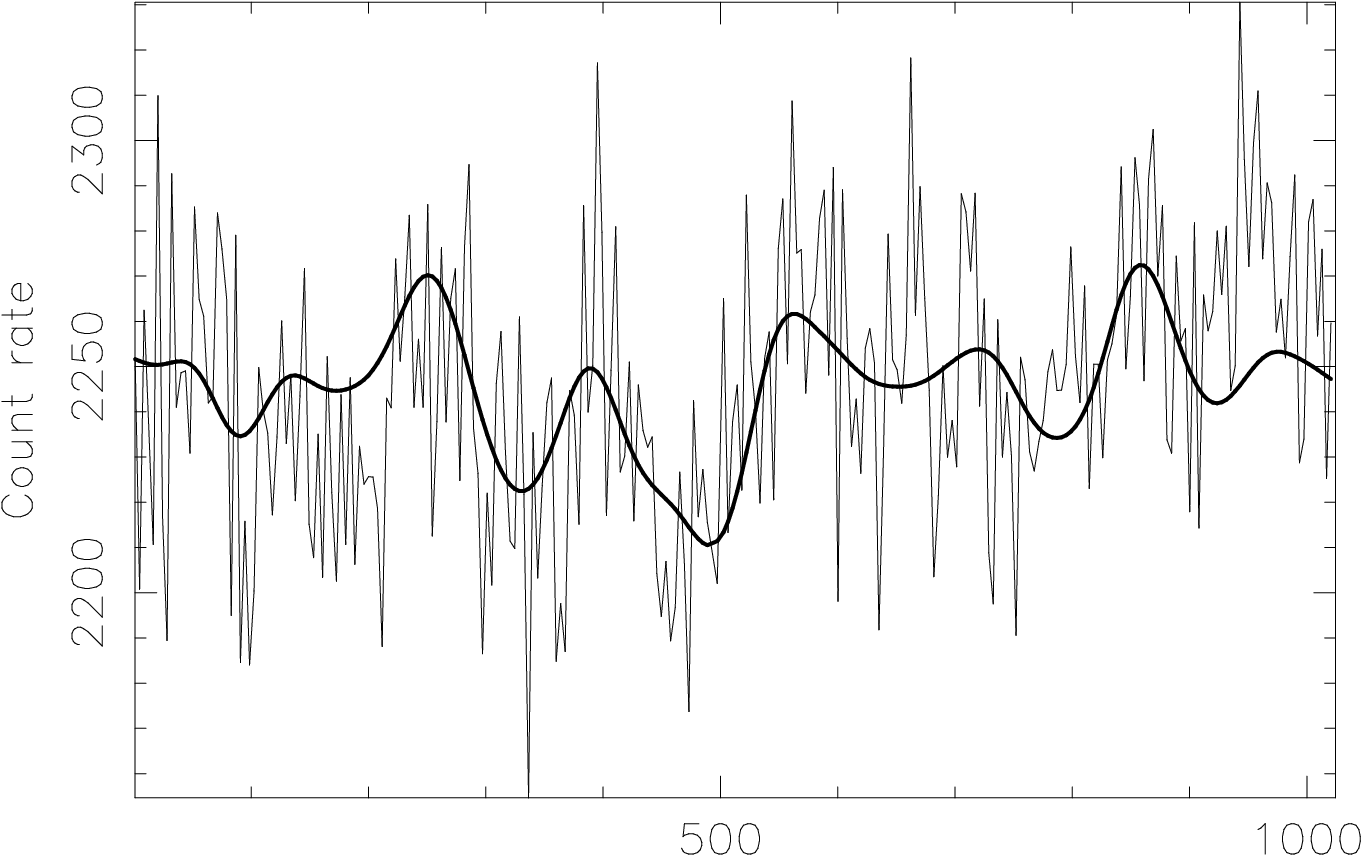}
\vspace*{2mm}
\epsfig{height=2.65in,angle=270,file=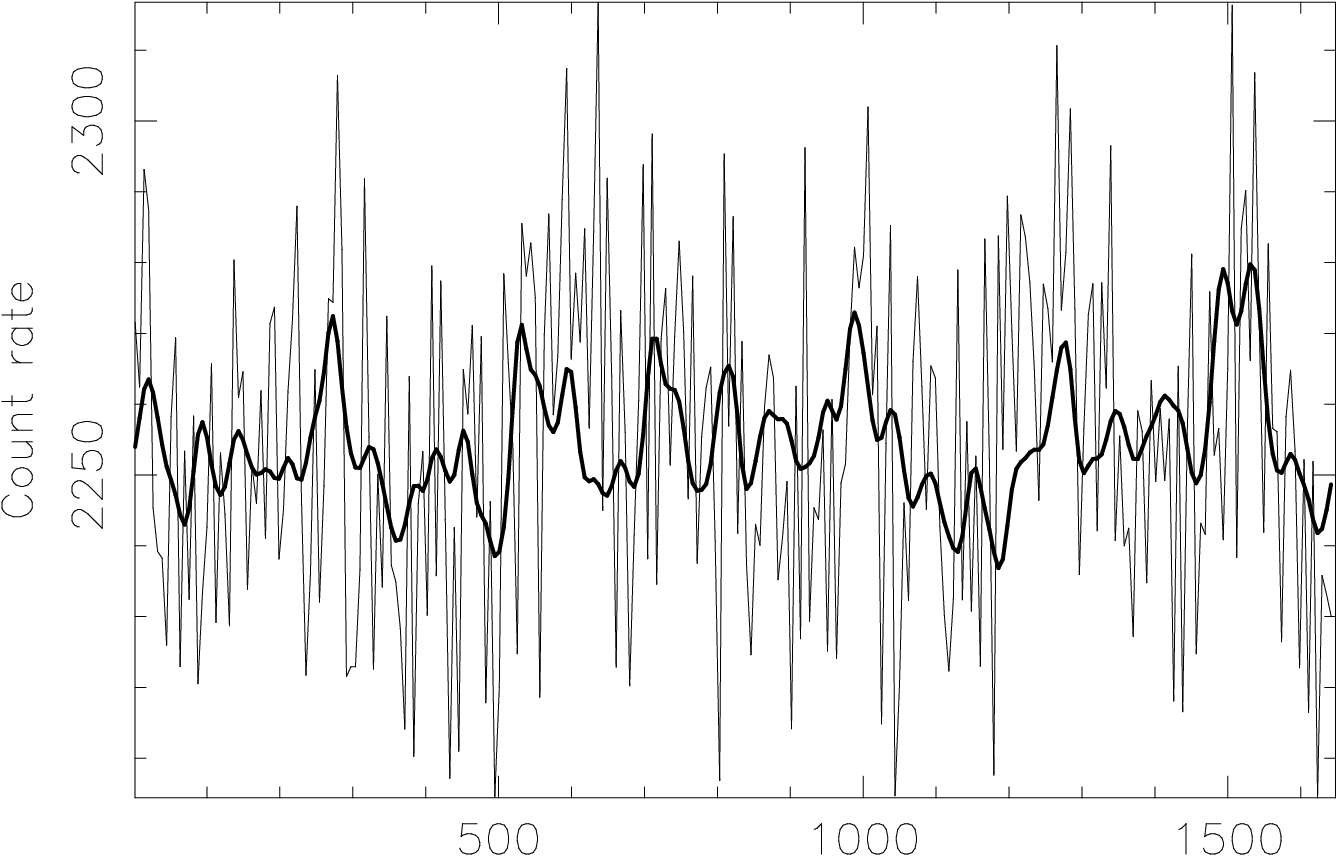}
\vspace*{2mm}
\epsfig{height=2.65in,angle=270,file=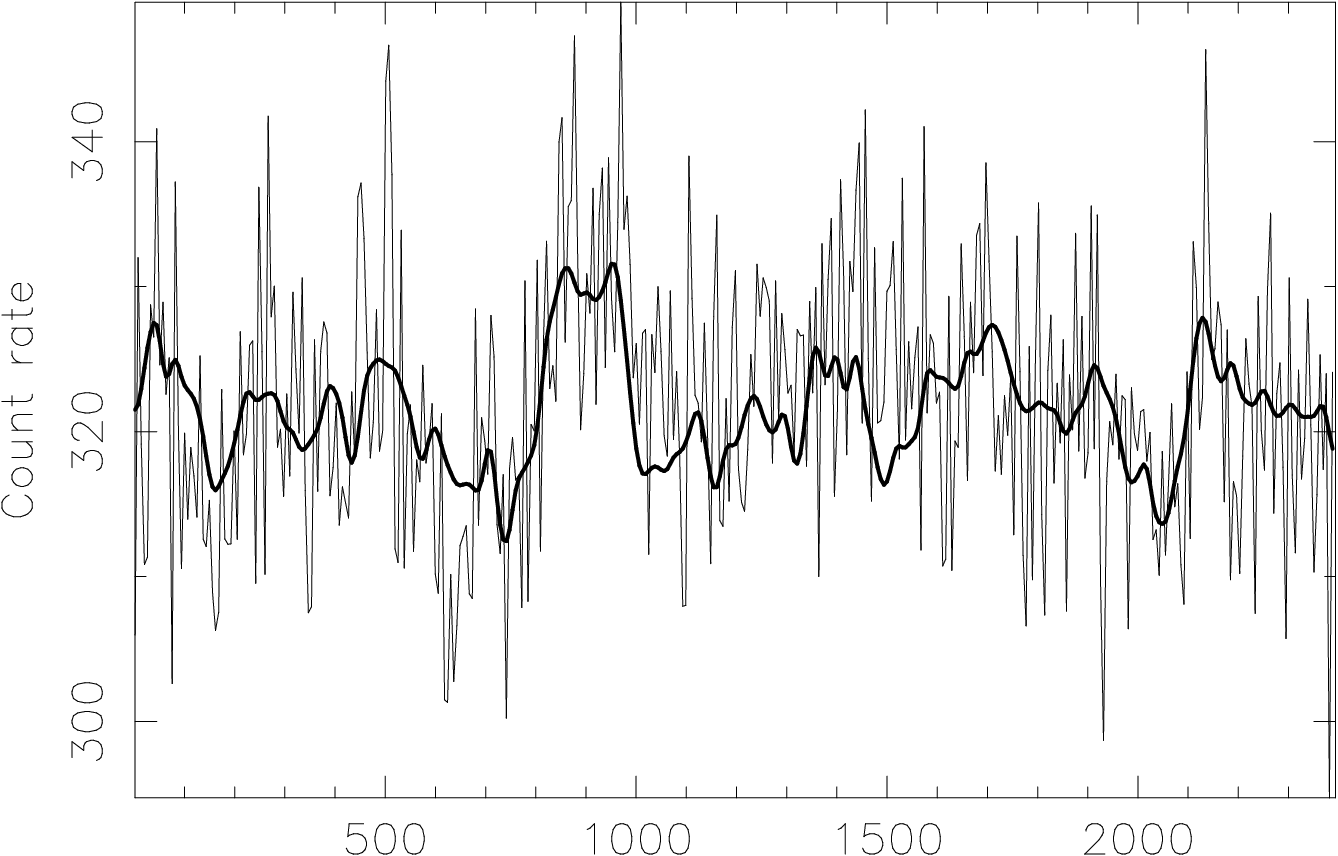}
\vspace*{2mm}
\epsfig{height=2.65in,angle=270,file=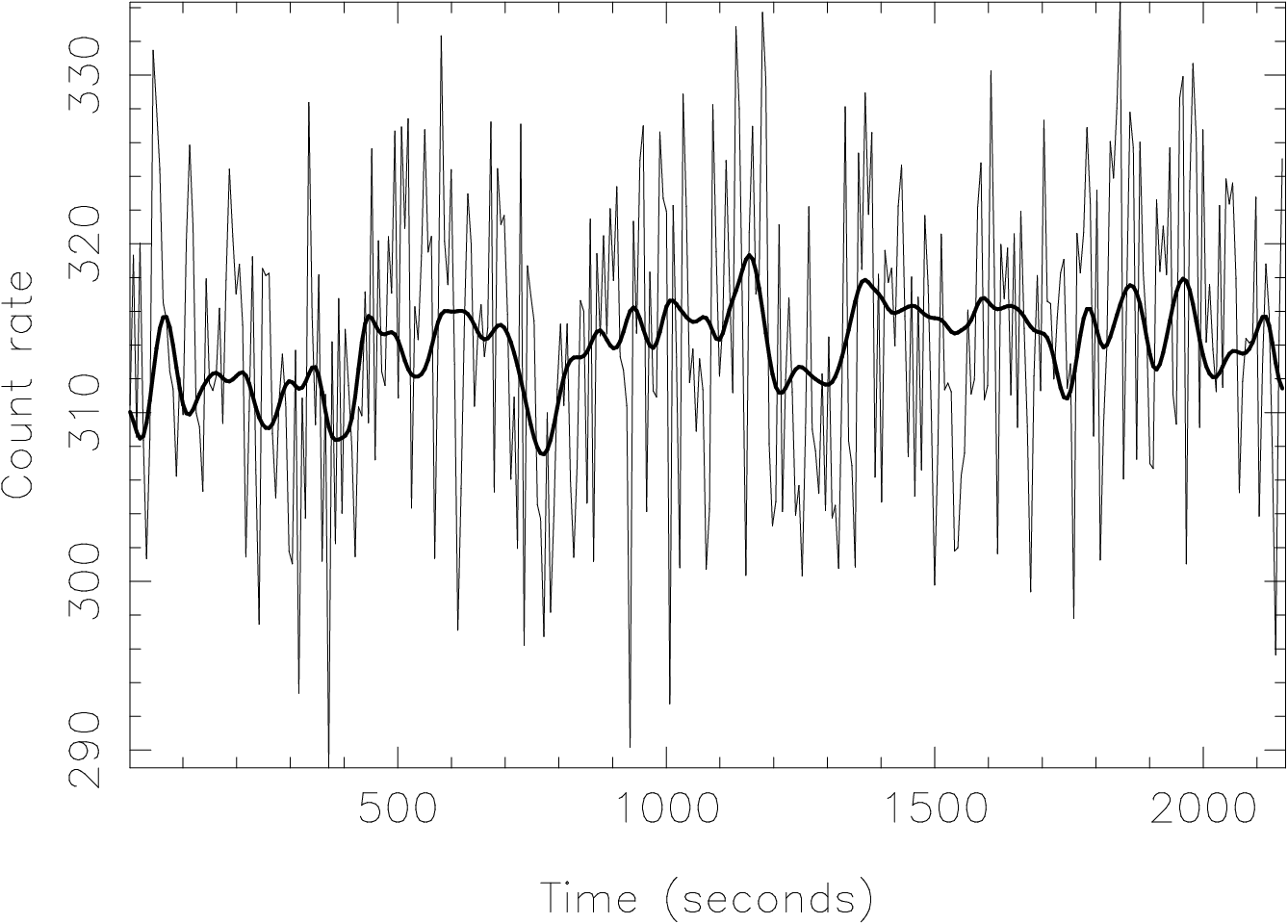}
\caption{Best-fit synthetic lightcurves for all 4 data sets from
Gaussian fitting, superimposed over actual \HST\ lightcurves.
The time axis is relative to the start of each exposure.  The
lightcurves have been rebinned by a factor of two for clarity.}
\label{GaussFitFigA}
\end{center}
\end{figure}

Figure~\ref{GaussFitFigA} shows the synthetic light curves from the
Gaussian superimposed over the 4 HST lightcurves. The
principal features of the \HST\ lightcurves are reproduced well in the
synthetic lightcurves.

\begin{figure*}
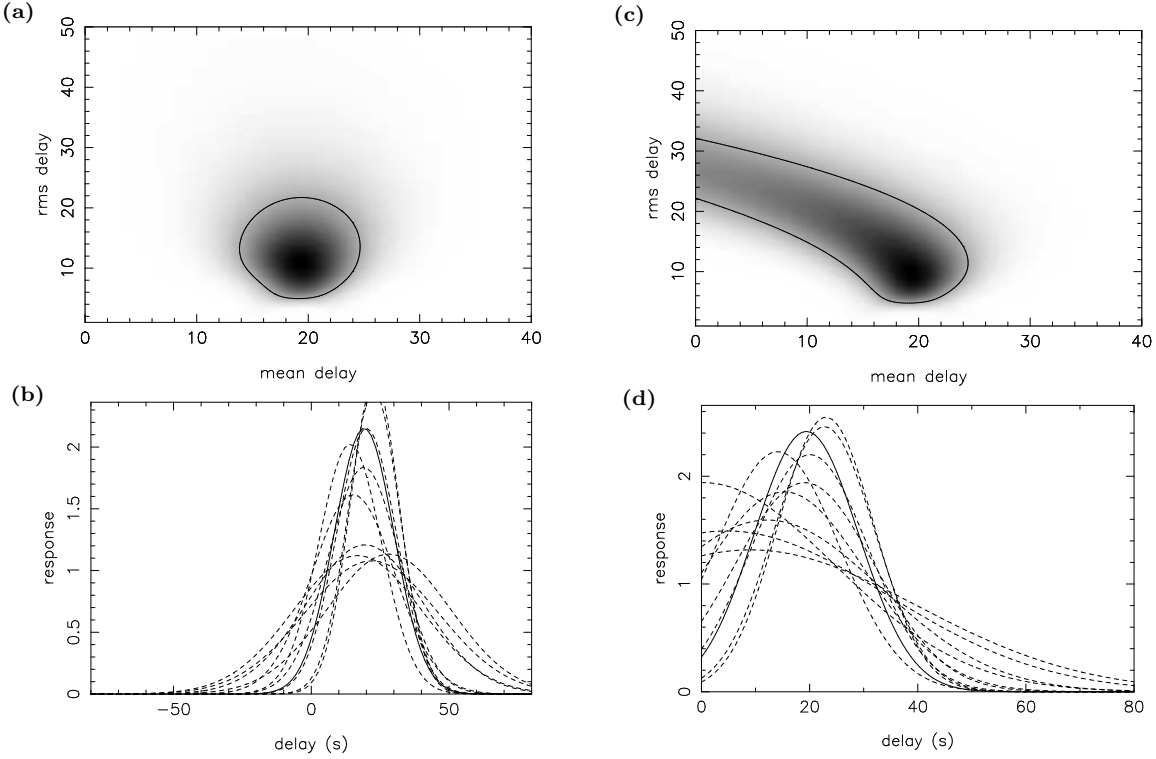

\begin{center}
\begin{minipage}{6.2in}
\begin{minipage}{3in}
\begin{center}
{\bf (a)}\epsfig{angle=270,width=2.65in,file=acausal.1.ps}
{\bf (b)}\epsfig{angle=270,width=2.55in,file=acausal.2.ps}
\end{center}
\end{minipage}
\hspace{\fill}
\begin{minipage}{3in}
\begin{center}
{\bf (c)}\epsfig{angle=270,width=2.65in,file=causal.1.ps}
{\bf (d)}\epsfig{angle=270,width=2.55in,file=causal.2.ps}
\end{center}
\end{minipage}
\end{minipage}
\caption{Top, acausal (left) and causal (right) $\chi^2$ surfaces for Gaussian
transfer function fitting to the \HST\ lightcurve 3. Bottom, best-fit
(solid line) and trial (dashed line) time-delay transfer functions to
HST Exp.\ 6.  The response is $\psi(\tau)$.}
\label{GaussFitFigB}
\end{center}
\end{figure*}

Figure~\ref{GaussFitFigB} shows the results of fitting the Gaussian
transfer functions to Exp.\ 6.  Panel (a) shows the constraints
imposed by the data on the mean and RMS delay.  The best fit for the
Gaussian fitting has $\chi^2_{min}/779 = 1.229 $. 
Here the 2-parameter 1-sigma confidence region is defined
by the contour $\chi^2 = \chi^2_{min} + 2.3$, and the greyscale
indicates relative probability.  Panel (b) uses Monte-Carlo error
propagation to indicate the range of uncertainty in the delay
distribution.  This plot shows 10 Gaussians selected at random with
the probability distribution indicated in panel (a).

The Gaussian model investigated above is acausal because it permits
\HST\ response preceding the X-ray driving.  We can construct a causal
model by truncating the negative delay tail of the
Gaussian. Panels (c) and (d) of Fig.~\ref{GaussFitFigB} illustrate the
results for the causal Gaussian fits in the same format as panels (a)
and (b) of the same figure.  The best fit for the causal Gaussian
fitting has $\chi^2_{min}/779 = 1.230 $. The
parameters $\tau_0$ and $\Delta \tau$ become anti-correlated because
the data to first order constrain the first and second moments of the
delay distribution.

\begin{table}
\begin{center}
\caption{Summary of results from parameterised Gaussian fitting to
         \HST\ lightcurves.  The rows give respectively the
         photometric phase, number of data points, reduced $\chi^2$,
         mean delay, RMS delay and integrated response for each light
         curve.  Uncertainties are based on 1-parameter, 1-sigma
         confidence regions.  The integrated response, $\Psi$, has
         been normalised by dividing by the mean of each \HST\ light
         curve.}
\label{GaussFitTable}
\begin{tabular}{lrrrr}
                       & Exp.\ 4 & Exp.\ 5 & Exp.\ 6 & Exp.\ 7 \\ 
\hline
$\phi_{\rm phot}$      & 0.36    & 0.39    & 0.42    & 0.44    \\
N                      & 528     & 541     & 782     & 704     \\ 
$\chi^{2}_{min}/(N-3)$ & 1.193   & 1.446   & 1.230   & 1.159   \\ 
\hline
$\tau_0$ (s) & $8.3\pm4.0$ 
             & $16.0^{+2.5}_{-2.2}$ 
             & $19.3\pm2.2$ 
             & $13.3^{+4.7}_{-4.5}$ \\
$\Delta\,\tau$ (s) & $20.0^{+6.8}_{-7.0}$ 
                   & $8.6^{+2.7}_{-2.3}$ 
                   & $10.8^{+3.7}_{-3.3}$ 
                   & $13.5^{+7.7}_{-4.7}$ \\
$\Psi/10^{-3}$ & $29\pm1$ 
               & $32\pm2$ 
               & $55^{+11}_{-7}$ 
               & $49\pm5$ \\ 
\hline
\end{tabular}
\end{center}
\end{table}

Table~\ref{GaussFitTable} summarises the results of fitting Gaussian
transfer functions to the 4 data segments.  For all 4 data segments
the mean and RMS delays are roughly consistent within the 1-parameter
1-sigma uncertainties. A weighted average yields $\overline{\tau_0} =
14.6\pm1.4$\,s, $\overline{\Delta\tau}=10.5\pm1.9$\,s.  The total
response, $\Psi$, is normalised using the mean count rate for the
individual light curve. This normalised total response appears to be
roughly constant for the two lightcurves from PRISM/RD, at a value of
$\Psi \sim 30\times10^{-3}$ and for the two lightcurves from PRISM/BL,
at a value of $\Psi \sim 50\times10^{-3}$. This difference between the
values of $\Psi$ shows that the variability is stronger at short
wavelengths, and suggests that the reprocessing therefore occurs in a
relatively hot part of the system.
%
%
\section{Discussion}
It is clear from our analysis that reprocessing with a mean time delay
of under 25\,s dominates.  This is the size of lag to be expected from
the accretion disc assuming established system parameters (Orosz \&
Bailyn 1997, van der Hooft et al.\ 1998).  Together with the
narrowness of the transfer function (RMS delay $\sim~10$\,s), this
means that disc reprocessing must dominate over the secondary star,
from which lags of greater than 40\,s are expected at this binary
phase ($\phi~\sim~0.40$). Even allowing for the maximum uncertainty in
system parameters estimated by van der Hooft et al.\ \shortcite{vdH98}, our
results are definitely inconsistent with the dominant source of
reprocessing being the companion star.  The response for the Gaussian
fits, when normalised to the count rate for the light curves are
consistent for the two PRISM/BL light curves and for the two PRISM/RD
ones.  The higher response, and hence reprocessing fraction, for the
PRISM/BL agrees with this analysis of the mean and RMS delays, showing
that the accretion disc is the most important region for reprocessing
of X-rays in \novasco.

Why is this the case?  During the activity observed by van der Hooft
et al.\ \shortcite{vdH97}, light curve analysis revealed that X-ray heating
of the companion star was important; if that was the case here, we
should see echoes originating from the companion star.  One
explanation for why we do not is that the X-ray absorbing material in
the disc may have a significant scale height above the mid-plane so
that the companion would effectively be shielded from direct X-ray
illumination, thus reducing the strength of reprocessing.  For this to
be the case, then the shielding material must rise to $H/R \sim 0.25$.

If the accretion disc in \novasco\ is significantly irradiated, we now
must ask why the long term optical light curve appears almost
anticorrelated with X-ray behaviour?  One possibility is that the
irradiating X-ray flux is being increasingly attenuated by a disc
corona.  As the optical depth of the Compton scattering corona
increases, we might naturally expect two consequences: firstly the
hard X-rays, believed to be produced by Comptonisation, would
increase.  Secondly, the irradiation of the outer disc would decrease,
as X-rays moving nearly parallel to the disc must penetrate a large
optical depth of scattering material.  We would thus expect that the
optical/UV and hard X-ray light curves might appear anticorrelated,
exactly as observed.  Indeed, a related scenario was proposed by
Mineshige \shortcite{M94} to explain why the onset of an optical
reflare in \novamus\ occured simultaneous to a decrease in the Thomson
optical depth deduced from X-ray spectra.

These observations represent a step forward in our understanding of
SXTs.  This is the first time that correlated optical--X-ray
variability has been detected from a transient source and supports
the picture, often assumed, that SXTs in outburst have irradiated
accretion discs similar to those in persistent LMXBs.
%
%
\section*{Acknowledgements}
RIH and KSO are supported by PPARC Research Studentships. RIH would
also like to acknowledge travel funding from the Nuffield Foundation.
Support for this work was provided by NASA through grant number
GO-6017-01-94A from the Space Telescope Science Institute, which is
operated by the Association of Universities for Research in Astronomy,
Incorporated, under NASA contract NAS5-26555 and also through contract
NAS5-32490 for the \RXTE\ project.  This work made use of the \RXTE\
Science Center at the NASA Goddard Space Flight Center and has
benefited from the NASA Astrophysics Data System Abstract Service.
Thanks to Ed Smith, Tony Keyes and Tony Roman at STScI for support.
%
%

%

\begin{thebibliography}{}
%
\bibitem[\protect\citename{Christensen et al.\ }1997]{C97}
        Christensen J. A., Welsh W. F., Evans I. N., Reinhart M.,
        Hayes J. J. E., 1997, CAL/FOS ISR 150, STScI
%
\bibitem[\protect\citename{Edelson \& Krolik }1988]{EK88}
        Edelson R. A., Krolik J. H., 1988, ApJ, 333, 646
%
\bibitem[\protect\citename{Gaskell \& Peterson }1987]{GP87}
        Gaskell C. M., Peterson B. M., 1987, ApJS, 65, 1
%
\bibitem[\protect\citename{Harmon et al.\ }1995]{Ha95}
        Harmon B. A., et al., 1995, Nat, 374, 703
%
\bibitem[\protect\citename{Hynes et al.\ }1998]{Hy98}
        Hynes R. I., et al., 1998, MNRAS, accepted
%
\bibitem[\protect\citename{Ilovaisky et al.\ }1980]{I80}
        Ilovaisky S. A., Chevalier C., White N. E., Mason K. O.
        Sanford P. W., Delvaille J. P., Schnopper H. W., 1980, 
        MNRAS, 191, 81
%
\bibitem[\protect\citename{Keyes et al.\ }1995]{K95}
        Keyes C. D., Koratkar A. P., Dahlem M., Hayes J., Christensen J.,
        Martin S., 1995, Faint Object Spectrograph Instrument
        Handbook, 6th Edn. 
        STScI, p. 52
%
\bibitem[\protect\citename{King \& Ritter }1998]{KR98}
        King A. R., Ritter H., 1998, MNRAS, 293, L42
%
\bibitem[\protect\citename{Middleditch \& Nelson }1976]{MN76}
        Middleditch J., Nelson J. E., 1976, ApJ, 208, 567
%
\bibitem[\protect\citename{Mineshige }1994]{M94}
        Mineshige S., 1994, ApJ, 431, L99
%
\bibitem[\protect\citename{Orosz \& Bailyn }1997]{OB97}
        Orosz J. A., Bailyn C. D., 1997, ApJ, 477, 876
%
\bibitem[\protect\citename{Peterson }1993]{P93}
        Peterson B. M., 1993, PASP, 105, 247 
%
\bibitem[\protect\citename{Petro et al.\ }1981]{P81}
	Petro L. D., Bradt H. V., Kelley R. L., Horne K., Gomer R.,
	1981, ApJ, 251, L7
%
\bibitem[\protect\citename{Remillard et al.\ }1996]{R96}
        Remillard R., Bradt H., Cui W., Levine A., Morgan E., Shirey B., 
        Smith D., 1996, IAU Circ.\ 6393
%
\bibitem[\protect\citename{Tanaka \& Shibazaki }1996]{TS96}
        Tanaka Y., Shibazaki N., 1996, ARA\&A, 34, 607
%
\bibitem[\protect\citename{van der Hooft et al.\ }1997]{vdH97}
	van der Hooft F. et al., 1997, MNRAS, 286, L43
%
\bibitem[\protect\citename{van der Hooft et al.\ }1998]{vdH98}
	van der Hooft F., Heemskerk M. H. M., Alberts F., van Paradijs
	J., 1998, A\&A, 329, 538
%
\bibitem[\protect\citename{White \& Peterson }1994]{WP94}
        White R. J., Peterson B. M., 1994, PASP, 106, 879
%
\end{thebibliography}
\end{document}